\documentclass[%
 reprint,
 showpacs,
 showkeys,
 floatfix,
 amsmath,amssymb,
 aps,
 pre
]{revtex4-1}

\usepackage{graphicx}% Include figure files
\usepackage{dcolumn}% Align table columns on decimal point
\usepackage{bm}% bold math
\title{Statics and dynamics of magnetocapillary bonds}

\begin{document}

\title{Statics and dynamics of magnetocapillary bonds}

\author{Guillaume Lagubeau}
 \altaffiliation[Also at ]{Departamento de F\`isica, Universidad de Santiago de Chile, Chile.}%Lines break automatically or can be forced with \\
\author{Galien Grosjean}%
 \email{ga.grosjean@ulg.ac.be}
\affiliation{%
GRASP - Physics Department, University of Li\`ege, B-4000 Li\`ege, Belgium. http://www.grasp-lab.org
}%
\author{Alexis Darras}
\affiliation{%
GRASP - Physics Department, University of Li\`ege, B-4000 Li\`ege, Belgium. http://www.grasp-lab.org
}%
\author{Geoffroy Lumay}
\affiliation{%
GRASP - Physics Department, University of Li\`ege, B-4000 Li\`ege, Belgium. http://www.grasp-lab.org
}%
\author{Maxime Hubert}
\affiliation{%
GRASP - Physics Department, University of Li\`ege, B-4000 Li\`ege, Belgium. http://www.grasp-lab.org
}%
\author{Nicolas Vandewalle}
\affiliation{%
GRASP - Physics Department, University of Li\`ege, B-4000 Li\`ege, Belgium. http://www.grasp-lab.org
}%

\date{\today}% It is always \today, today,
             
\begin{abstract}
When ferromagnetic particles are suspended at an interface under magnetic fields, dipole-dipole interactions compete with capillary attraction.
This combination of forces has recently given promising results towards controllable self-assemblies, as well as low Reynolds swimming systems.
The elementary unit of these assemblies is a pair of particles.
Although equilibrium properties of this interaction are well described, dynamics remain unclear.
In this letter, the properties of magnetocapillary bonds are determined by probing them with magnetic perturbations.
Two deformation modes are evidenced and discussed.
These modes exhibit resonances whose frequencies can be detuned to generate non-reciprocal motion.
A model is proposed which can become the basis for elaborate collective behaviours.
\end{abstract}

\pacs{81.16.Dn, 68.03.Cd, 47.63.mf}% PACS, the Physics and Astronomy
\keywords{Self-assembly, surface tension, low Reynolds number}%Use showkeys class option if keyword
\maketitle

%\tableofcontents

%%%%%%%%%%%%%%%%%%%%%%%%%%%%%%%%%%%%%%%%%%%%%%%%%%%%%%%%%%%%%%%%%%%%%%%%%%%%%%%%%%%%%%%%%%%%%%%%%%%%%%%%%%%%%%%%%%%%%%%%%%%
\section{Introduction}

The deformation of a liquid surface by identical floating particles~\cite{vella2015} makes them attract each other~\cite{kralchevsky2001,vella2005}.
A practical way to avoid clustering consists in using particles possessing a large magnetic moment $\vec{m}$~\cite{golosovsky1999,vandewalle2012}.
If $\vec{m}$ is perpendicular to the interface, the dipole-dipole interaction is repulsive~\cite{jackson1999} and opposes the capillary attraction.
The combination of a repulsive magnetic interaction with the attractive capillary interaction creates a pair potential possessing a minimum.
Two particles will then settle at an equilibrium distance creating a `magnetocapillary bond' which is the building block for larger stable structures~\cite{golosovsky1999,vandewalle2013}.
Indeed the energy scale associated with this magnetocapillary interaction can be much larger than the thermal energy for a wide range of submillimetre-sized particles. 
Particle size is bounded from below by the capillary attraction, which decreases as particles become smaller.
At room temperature, the size at which thermal energy becomes strong enough to break free from the capillary interaction is $3.4\;\mathrm{\mu m}$ in diameter, as can be estimated from the potential given in equation~\ref{Uc}.
However, one could generate a stronger capillary attraction by using confined geometries~\cite{kralchevsky1994, ershov2013} or other body forces than gravity~\cite{vella2015}.
To allow remote tuning of the bond properties, it has been proposed to use particles with large magnetic susceptibility $\chi$ in an external magnetic induction $\vec{B}$~\cite{chinomona2015}.
In this case $\vec{m}=\chi V \vec{B} / \mu_0$ (with V the volume of the bead), meaning that varying the magnetic induction allows to control the strength and direction of the magnetocapillary interaction~\cite{vandewalle2012}.
Adding an oscillating horizontal field deforms these aggregates~\cite{chinomona2015} which are observed to swim at low Reynolds number~\cite{lumay2013} by spontaneously changing their shape in a non-reciprocal way and, what is more, swimming direction can be remotely controlled~\cite{grosjean2015}.

\begin{figure}
\begin{center}
\includegraphics[width=85mm]{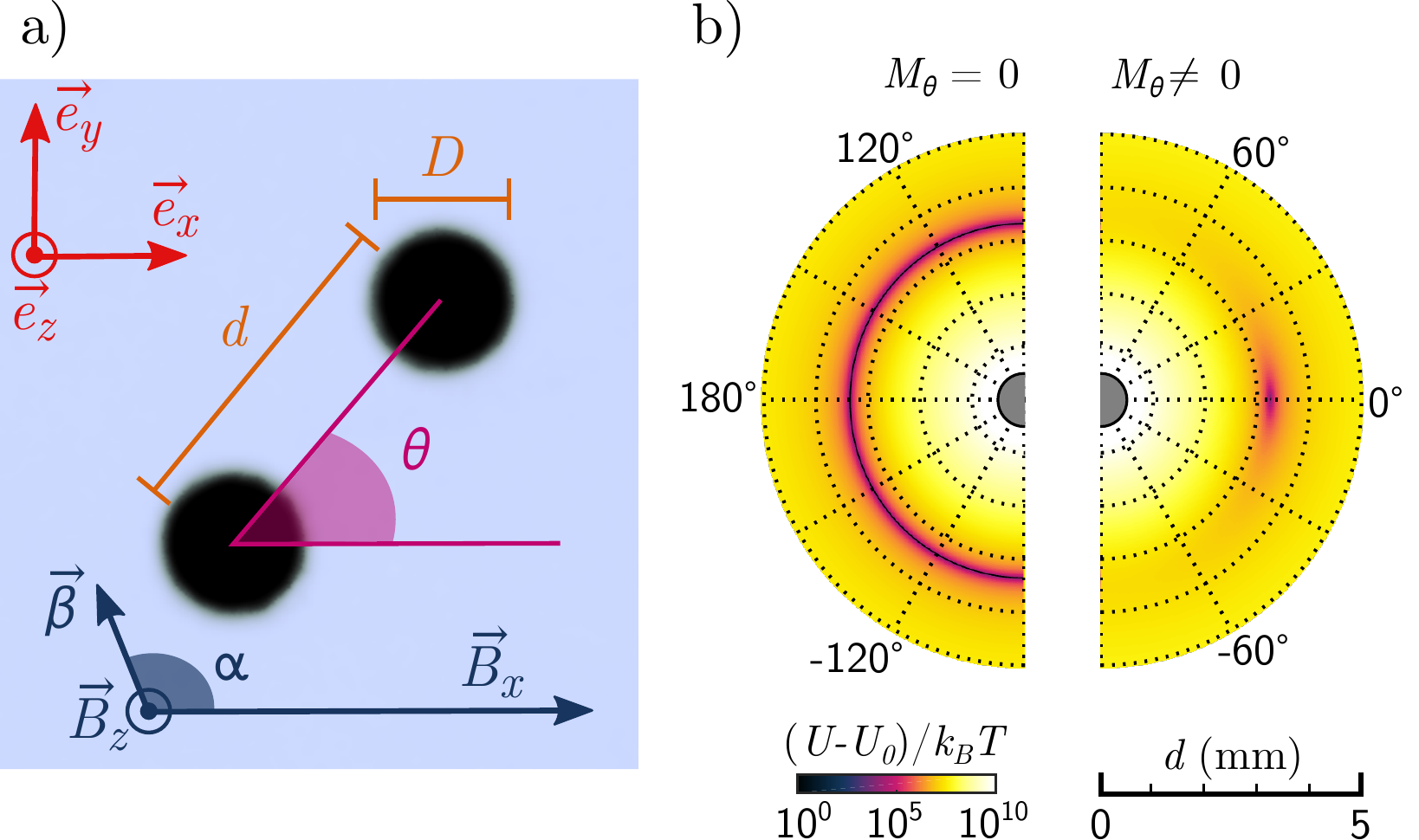}
\caption{
\textbf{Pair potential.} -- 
\textbf{a)~Notations.}
Two beads are placed at a water surface and submitted to a constant induction $\vec{B}=B_x \vec{e_x}+B_z\vec{e_z}$ as well as a small oscillating induction $\vec{\beta}$.
Distance $d$ and angle $\theta$ characterise the system.
\textbf{b)~Magnetocapillary potential.}
Interaction potential $U = U_{c}+U_{m}$ in polar coordinates $(d,\theta)$, as given by eq.~(\ref{U}), for typical parameters of the experiment. 
Darker zones indicate potential wells.
On the left half, vertical induction $\vec{B_z}$ generates an isotropic repulsion when $M_\theta = 0$, which leads to a finite equilibrium distance $d_{eq} \approx 2.5$ mm. 
On the right half, cylindrical symmetry is broken when $M_\theta \neq 0$.
$M_\theta$ is the orthoradial magnetocapillary number introduced in eq.~(\ref{U}). It is proportional to $B_x^2$.}
\label{FIG1}
\end{center}
\end{figure}

The aim of the present letter is to study experimentally the fundamentals of the magnetocapillary bonds and their vibration modes.
First, we present their potential energy and their static equilibrium distances. 
Then, their two normal vibrations (stretching and swinging) are evidenced experimentally as well as modelled in a perturbative way.
Finally, we show that the detuning of their resonance frequencies can be exploited for creating non-reciprocal cycles at low Reynolds number.
Although particles of a few hundred microns in diameter were used, if the system is downscaled, resonance of the swinging mode can still be achieved.
This means that similar dynamics could be reproduced with smaller, colloidal particles~\cite{euan2015}.

%%%%%%%%%%%%%%%%%%%%%%%%%%%%%%%%%%%%%%%%%%%%%%%%%%%%%%%%%%%%%%%%%%%%%%%%%%%%%%%%%%%%%%%%%%%%%%%%%%%%%%%%%%%%%%%%%%%%%%%%%%%
\section{Magnetocapillary pair potential}

We consider two floating beads of diameter $D$ possessing identical magnetic moments $\vec{m}=m_x\vec{e_x} +m_z\vec{e_z}$, $\vec{e_x}$ and $\vec{e_z}$ being the horizontal and vertical unitary vectors, respectively, as defined in fig.~\ref{FIG1}~a).
Their center-to-center vector is denoted $\vec{d}=d \vec{e_r}$, and the orientation of the pair relatively to $\vec{e_x}$ is $\theta$.
Their interaction can be modelled by a pair potential $U$ composed of the sum of an attractive term due to capillarity $U_c$ and a magnetic dipole-dipole $U_m$ term that can be either attractive or repulsive depending on $\theta$~\cite{chinomona2015, grosjean2015}.

A reasoning similar to~\cite{kralchevsky2001} will be used to express the capillary interaction energy $U_c$.
Solving Laplace equation for a single sphere gives interfacial shape $\zeta \left( d \right) \propto K_0 \left( d / l_c \right)$, where $l_c=\sqrt{\gamma/\rho g}$ is the capillary length and $K_0$ is the modified Bessel function of the second kind and order 0.
If $d\gg D$, the superposition approximation can be used to express $U_c$, meaning that the total interfacial deformation is the sum of the independant deformations caused by each particle~\cite{nicolson1949}. 
As will be shown in the discussion of figure~\ref{FIG4}, this approximation is in good agreement with the experiment for $d>2D$. We find
\begin{equation}
U_c=-\Gamma K_0(d/l_c)
\label{Uc}
\end{equation}
with $\Gamma=2\pi \gamma q^2$ being a typical capillary energy scale.
The capillary charge of a particle $q$ is a length that characterises the interface deformation needed to compensate the buoyancy of the particle. It depends explicitly on the bead volume, the density of the bead, the density of the liquid, the capillary length and the wetting contact angle~\cite{kralchevsky2001,vella2005}.
If none of the fluids has strong magnetic properties, the magnetic dipole-dipole potential is
\begin{equation}
U_m=\frac{\mu_0 \left[ m_z^2+m_x^2 (1-3\cos^2 \theta) \right]}{4\pi d^3}
\end{equation}
with $\mu_0$ the void permeability.
If the distances are nondimensionalised by the capillary length ($\tilde{d}=d/l_c$), the shape of pair potential $U = U_c + U_m$ is determined by two dimensionless magnetocapillary numbers $M_d$ and $M_\theta$ such that
\begin{equation}
U=\Gamma \left[ -K_0(\tilde{d})+ \frac{M_d}{\tilde{d}^3}+\frac{M_\theta}{\tilde{d}^3} \sin^2\theta \right]
\label{U}
\end{equation}
where $M_d=\mu_0 (m_z^2-2m_x^2) /4\pi \Gamma l_c^3$ is the radial magnetocapillary number and $M_\theta=3\mu_0 m_x^2 /4\pi \Gamma l_c^3$ the orthoradial magnetocapillary number.

While the magnetic terms of $U$ are dominating at both short and long distances, it has a stable minimum provided that  $M_d<1.11$.
Thus, it forms a potential well around $\theta=0$ and the equilibrium distance ${d}_{eq}$.
It can be shown that $\tilde{d}_{eq} = d_{eq}/l_c$ depends only on $M_d$ following the equation 
\begin{equation}
\tilde{d}_{eq}^4 K_1(\tilde{d}_{eq})=3 M_d.
\label{deq}
\end{equation}
If $M_\theta=0$, the potential is axisymmetric and every orientation $\theta$ is equivalent for the pair. 
Figure~\ref{FIG1}~b) shows the interaction potential $U$ in polar coordinates, with one bead at the axis origin, for typical values of the experiment. 
Because the potential is mirror symmetric around the $90^\circ$ axis, the two halves of the graph are chosen to represent two cases.
Case $M_\theta=0$ is on the left, with a minimum of energy at a given distance ${d}_{eq}$.
On the right, the addition of a horizontal induction $B_x$ of order $10^{-3}$ T ($M_\theta \neq 0$) breaks axisymmetry and creates an equilibrium orientation at $\theta=0$.
Typical values for the experiment are $B_z \approx 5 \times 10^{-3}$ T and $B_x \approx 2 \times 10^{-3}$ T, which gives a ratio $M_\theta / M_d \approx 0.8$.

Using a perturbative analysis of $U$, the stiffness of the potential well can be obtained for its two principal direction (radial $k_d$ and orthoradial $k_\theta$):
\begin{equation}
k_d =\frac{\Gamma} {l_c^2\tilde{d}} \left[ 6K_1(\tilde{d})-2\tilde{d} K_0(\tilde{d}) \right]
\end{equation}
\begin{equation}
k_\theta =4\frac{\Gamma M_\theta^2}{l_c^2\tilde{d}^5}.
\end{equation}
In the case of two identical beads of mass $W_1=W_2=W$, the reduced mass is $W/2$.
Two frequencies can thus be associated with the magnetocapillary well: a radial frequency
\begin{equation}
\omega_d = 2\pi f_d = \sqrt{2k_d / W} \approx
\frac{2}{l_c} \sqrt{\frac{U_c (\tilde{d}_{eq})}{W} + \frac{9 M_d \Gamma}{W \tilde{d}^5_{eq}}},
\label{wd}
\end{equation}
assuming $d \gg D$, that corresponds to a stretching mode and an orthoradial frequency
\begin{equation}
\omega_\theta = 2\pi f_\theta = \sqrt{2k_\theta / W} =
\frac{2 M_\theta}{l_c} \sqrt{\frac{2 \Gamma}{W \tilde{d}^5_{eq}}}
\label{wtheta}
\end{equation}
that corresponds to a swinging mode.
At low Reynolds, the attenuation is due to the viscosity of the ambient fluid, so that attenuation time is
\begin{equation}
\tau = \frac{W}{C} \sim \frac{W}{\eta D}
\end{equation}
where $C$ is the damping coefficient.
We have $C_{Stokes} = 3\pi\eta D$ in the case of a fully immersed sphere~\cite{petkov1995,danov2000}.
If $f_d \tau \ge 1$ or $f_\theta \tau \ge 1$, we can force harmonic resonances, even in the Stokes regime.
This is of interest as frequency can thus be changed to adjust phase.

%%%%%%%%%%%%%%%%%%%%%%%%%%%%%%%%%%%%%%%%%%%%%%%%%%%%%%%%%%%%%%%%%%%%%%%%%%%%%%%%%%%%%%%%%%%%%%%%%%%%%%%%%%%%%%%%%%%%%%%%%%%
\section{Experimental setup}

\begin{figure}
\begin{center}
\includegraphics[width=85mm]{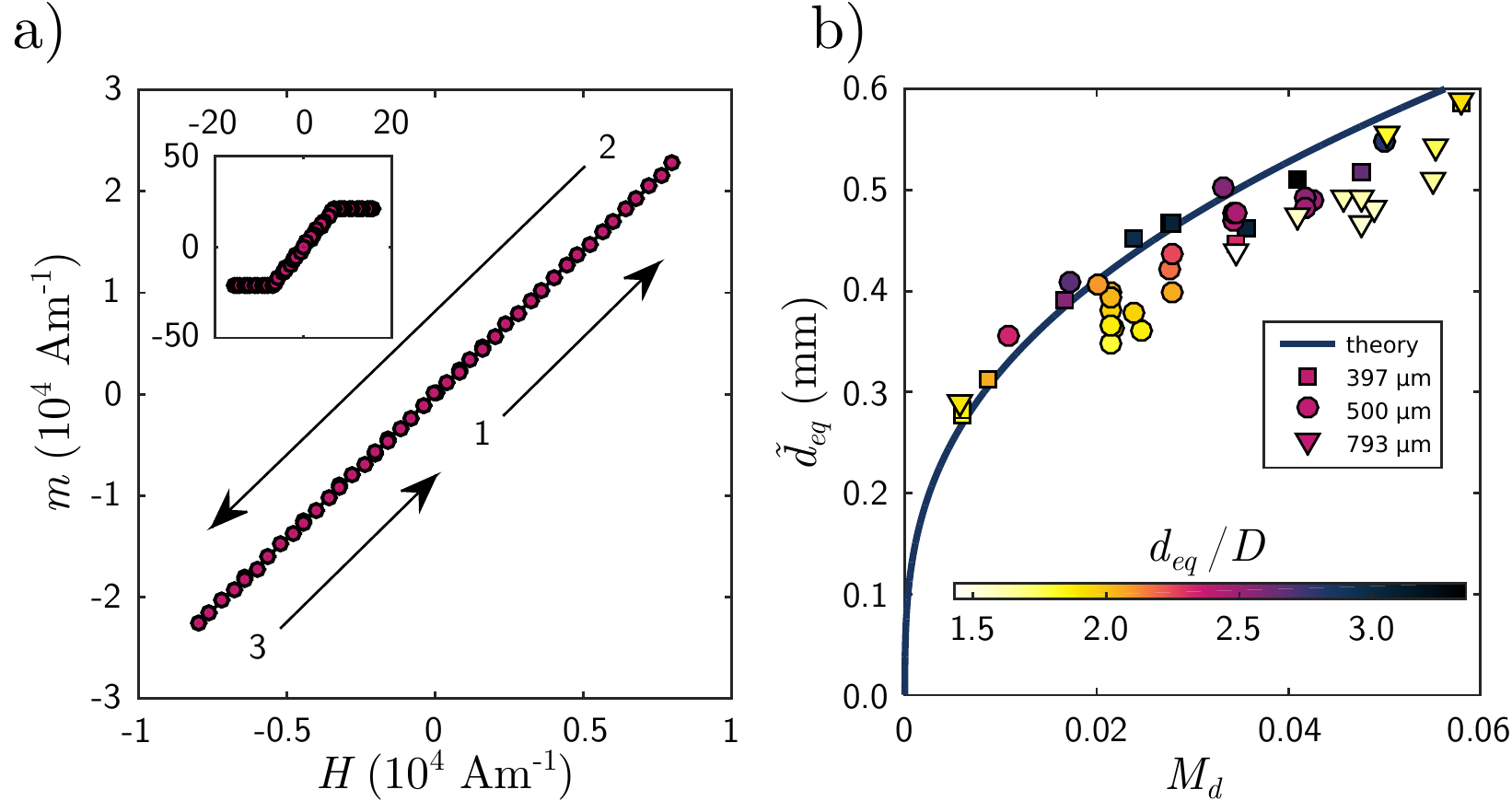}
\caption{
\textbf{Characterisation of the system.} -- 
\textbf{a)~Beads magnetisation.}
Magnetic field $H$ induces a magnetisation $m$ in a 500 $\mathrm{\mu m}$ bead with negligible hysteretic behaviour.
$H$ is successively increased from 0 (1), decreased (2) and increased back to 0 (3).
Values of $H$ are typical of the experiment.
Inset shows a larger cycle up to saturation magnetisation, with same units.
\textbf{b)~Equilibrium distance.}
The dimensionless equilibrium distance $\tilde{d}_{eq}$ as a function of the radial magnetocapillary number $M_d$. 
The colourmap shows distance $d_{eq}$ over diameter $D$.
Solid line corresponds to theoretical prediction without adjustable parameter, obtained by numerically solving eq.~(\ref{deq}). Symbols represent different bead diameters.
}
\label{FIG2}
\end{center}
\end{figure}

In order to study the vibrations of magnetocapillary bonds, we will use the following experimental setup: a large Petri dish is filled with water and placed at the center of a three-axis, Earth's field compensating Helmholtz coil system.
Two identical submillimetric chrome steel beads (diameter $D=397$, $500$ or $793\; \mathrm{\mu m}$, density $\rho = 7830 \;\mathrm{kg/m^3}$, relative magnetic susceptibility $\chi > 300$) are disposed at the water-air interface. 
The capillary charges of the beads, calculated for a contact angle of $90^\circ$, are $6$, $12$ and $45 \;\mathrm{\mu m}$, for the respective diameters $397$, $500$ and $793 \;\mathrm{\mu m}$.
In the presence of a constant magnetic induction $\vec{B}=B_x \vec{e_x}+B_z\vec{e_z}$, the beads settle at an equilibrium distance $d_{eq}$.
For vibration experiments, a small oscillating induction $\vec{\beta}=\beta (t) (\cos\alpha\,\vec{e_x}+\sin\alpha\,\vec{e_y})$ is added.
Angle $\alpha$ can be varied to produce different oscillating behaviours.
The kinematics of the system is fully described by the beads center-to-center distance $d$ and its rotation angle $\theta$ as defined in fig.~\ref{FIG1}~a).

For obtaining the desired magnetic properties, we chose chrome steel spheres (alloy AISI 52100).
The bulk material is ferromagnetic and has a large susceptibility $\chi$, which depends on applied magnetic field $H = B/\mu_0$, such that $\chi (H) > 300$~\cite{blazek2014}.
However, a finite ferromagnetic body in a magnetic field produces a demagnetisation effect.
An effective susceptibility $\chi_{\text{eff}}$ takes this effect into account, which for a sphere has the form $\chi_{\text{eff}} = \chi / \left( 1 + \chi / 3 \right)$ ~\cite{osborn1945}.
For $\chi \gg 3 $, the sphere will thus behave like an isotropic superparamagnetic particle of susceptibility $\chi_{\text{eff}} = 3$.
In fig.~\ref{FIG2}~a), we plot the experimental magnetisation curve $m(H)$.
Indeed, no hysteresis is observed, and the slope gives $\chi_{\text{eff}} \simeq 3$ as expected.
A small magnetisation remains at zero field, of the order of 100 $\mathrm{Am^{-1}}$, which is negligible for the fields used herein (typically 5000 $\mathrm{Am^{-1}}$).

%%%%%%%%%%%%%%%%%%%%%%%%%%%%%%%%%%%%%%%%%%%%%%%%%%%%%%%%%%%%%%%%%%%%%%%%%%%%%%%%%%%%%%%%%%%%%%%%%%%%%%%%%%%%%%%%%%%%%%%%%%%
\section{Results}

Equilibrium distance $d_{eq}$ has been measured for 3 diameters $D$ and various values of radial magnetocapillary number $M_d$, changed by varying both $B_x$ and $B_z$.
As seen in fig.~\ref{FIG2}~b), the equilibrium distance calculated from the interaction potential, as defined in eq.~(\ref{deq}), is in good agreement with experimental values for large distance $d_{eq}$ compared to diameter $D$.
Indeed, for $d_{eq}/D \lesssim 2$ (light-coloured points in the graph), the superposition approximation no longer holds in the determination of capillary potential $U_{c}$~\cite{he2013}. In this case, the attraction is stronger.

\begin{figure}
\begin{center}
\includegraphics[width=85mm]{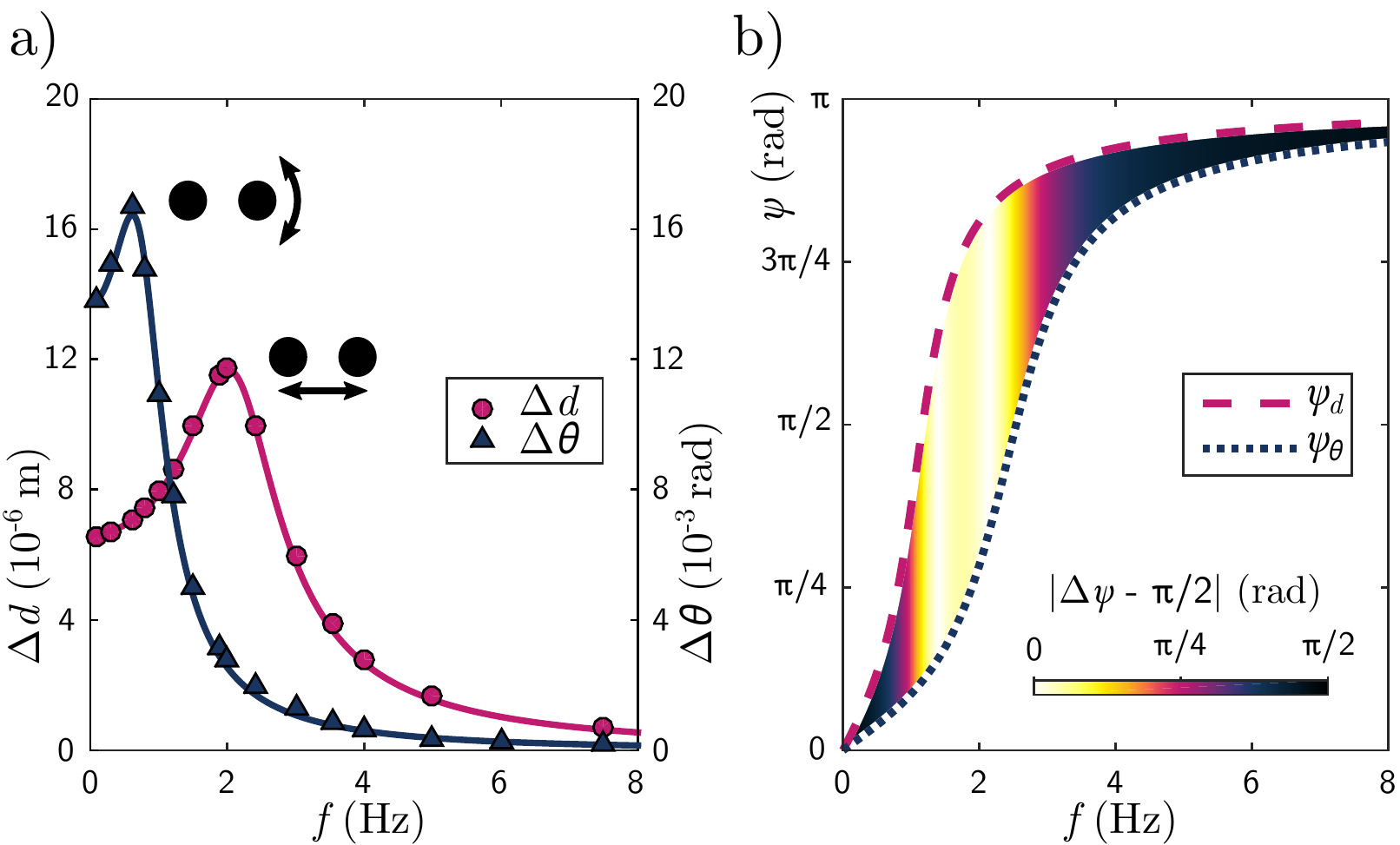}
\caption{
\textbf{Resonance spectrum.} -- 
\textbf{a)~Amplitude.}
Frequency response of distance $d$ and angle $\theta$ for typical values of the experimental parameters ($D = 500 \;\mathrm{\mu m}$, $M_\theta / M_d \approx 0.8$).
Radial oscillation of amplitude $\Delta d$ happens for $\alpha = 0^\circ$, while orthoradial oscillation of amplitude $\Delta \theta$ happens for $\alpha = 90^\circ$.
Error bars are smaller than the symbols, with a typical error of $0.2\;\mathrm{\mu m}$ on the bead center.
Data have been fitted with a resonance curve, represented by solid lines.
\textbf{b)~Phase.}
Theoretical prediction for the phase in both cases.
Colourmap indicates phase difference $\Delta \psi$.
}
\label{FIG3}
\end{center}
\end{figure}

In order to characterise the properties of the magnetocapillary well, a horizontal oscillating induction $\vec{\beta} (t)$ with magnitude $\beta (t) = \beta \sin ( 2\pi f t )$ and angle $\alpha$ is added.
A small oscillation amplitude $\beta \ll |\vec{B}_{x}|$ is chosen, so that the system remains close to its equilibrium configuration.
Pertubative analysis gives two oscillation modes in radial and orthoradial directions, that we will excite independently.
Two values of angle $\alpha$ corresponding to both directions will thus be studied: $\alpha = 0^\circ$ (radial) and $\alpha = 90^\circ$ (orthoradial).

\begin{figure}
\begin{center}
\includegraphics[width=85mm]{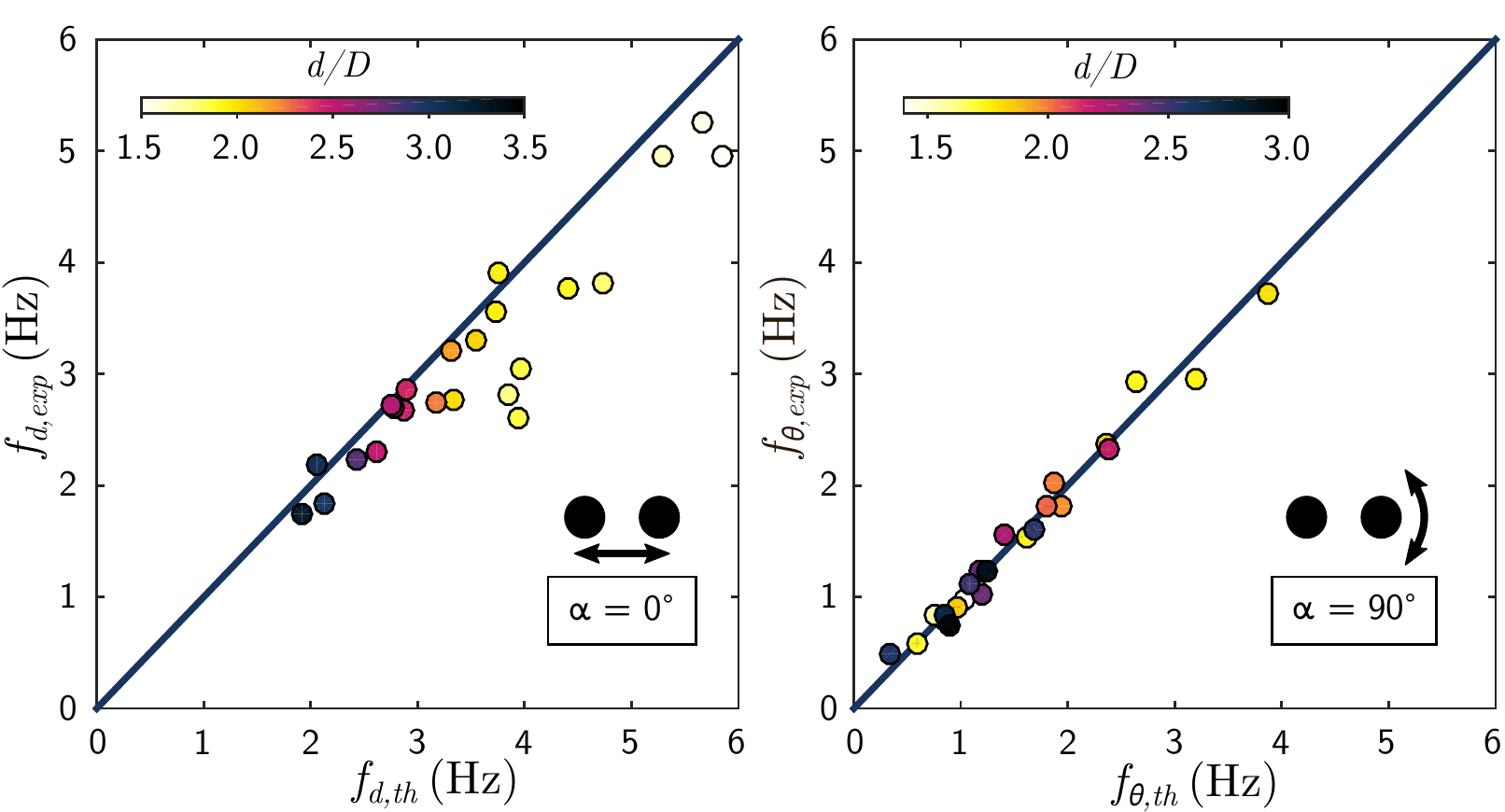}
\caption{
\textbf{Resonance frequencies.} -- 
Comparison between measured resonance frequency $f_{exp}$ and theoretical prediction $f_{th}$, for several values of induction fields $B_{x}$, $B_{z}$ and $\beta$ as well as several bead diameters.
Two orientations $\alpha$ of $\vec{\beta}$ are explored which correspond to radial and angular oscillations, respectively $\alpha = 0^\circ$ on the left and $\alpha = 90^\circ$ on the right.
Theory and experiment agree well, with the exception of distances $d/D \lesssim 2$ in case $\alpha = 0^\circ$, corresponding to light-coloured points in the left graph.
}
\label{FIG4}
\end{center}
\end{figure}

In fig.~\ref{FIG3}~a), oscillation amplitudes of distance $d$ and angle $\theta$ are represented as a function of frequency.
Each point was obtained by identifying the geometric center of each bead, averaged over 10 oscillation periods, in order to reach sub-pixel precision.
Two independent oscillation modes can indeed be excited: a radial mode of amplitude $\Delta d$ for $\alpha = 0^\circ$, for which $d$ oscillates around $d_{eq}$ and $\theta = 0$ (stretching mode), and an orthoradial mode of amplitude $\Delta \theta$ for $\alpha = 90^\circ$, called swinging mode, for which $\theta$ oscillates around 0 and $d=d_{eq}$.
Both modes exhibit a resonance, with distinct resonance frequencies $f_d > f_\theta$ in this example.

Steady-state solution of a forced damped harmonic oscillator is usually defined as the real part of a complex number $z (t) = R (\omega) e^{i\omega t}$ with function $R (\omega) = A (\omega) e^{-i \psi (\omega)}$.
If $F_0 /W$ is the driving force per unit mass, we have the following expressions for amplitude $A (\omega)$ and phase $\psi (\omega)$, respectively
\begin{equation}
A (\omega) = \frac{F_0 / W}{\sqrt{(\omega_0^2-\omega^2)^2 + 4\omega^2/\tau^2}}
\end{equation}
and
\begin{equation}
\psi (\omega) = \arctan \left( \frac{2 \omega}{\tau (\omega_0^2 - \omega^2)} \right),
\end{equation}
with angular frequency $\omega=2\pi f$, resonance frequency $\omega_0=2\pi f_0$ and damping time $\tau$.
Data in fig.~\ref{FIG3}~a) are very well fitted by function $A (\omega)$, with fit parameters $F_0/W$, $\omega_0$ and $\tau$.
The resonance curves correspond to underdamped harmonic oscillators as the quality factor $Q=\omega_0 \tau/2$ ranged between 0.6 and 3.8.
The resonance frequencies in all our experiments ranged between 0.5 and 5.2 Hz.
Finally, the time scale $\tau$ given by the fit corresponds to an average viscous damping coefficient $C$ that grows linearly with bead diameter $D$ and is close to a Stokes damping $C_{Stokes} = 3\pi\eta D$, such that $C/C_{Stokes} = 0.86 \pm 0.04$.
This seems consistent with the viscous damping of a partially immersed sphere~\cite{petkov1995,danov2000}.
The dispersion of the values can be due to the interaction between the spheres, as well as slight variations of the wetting conditions.
Note that this average value of $C$ excludes radial oscillations for which $d/D \lesssim 2$, which are discussed below.
Figure~\ref{FIG3}~b) represents the corresponding phases $\psi$.
Because radial and orthoradial resonance frequencies are distinct, a phase difference $\Delta \psi$ can arise between modes.
Close to resonance frequencies, we have $\Delta\psi \approx \pi / 2$.

Radial and orthoradial resonance frequencies, respectively $f_d$ and $f_\theta$, are determined for different sets of parameters $B_{x}$, $B_z$, $\beta$ and $D$.
Figure~\ref{FIG4} compares experimental values $f_{d,\;exp}$ and $f_{\theta,\;exp}$ with theoretical predictions $f_{d,\;th}$ and $f_{\theta,\;th}$ as defined in eqs.~(\ref{wd}) and~(\ref{wtheta}).
Case $\alpha = 0^\circ$ shows good agreement for large values of $d/D$, but theory gives an overestimation of frequency $f_d$ for distances $d/D \lesssim 2$.
This evidences the limits of validation of the superposition approximation in capillary potential $U_{cap}$.
In case $\alpha = 90^\circ$, by contrast, theory and experiment agree well.
Indeed, linear regression gives $f_{d,\;exp} \approx f_{d,\;th}$ down to about 1\%.
Because distance $d$ is constant in the ortho\-radial case, capillarity does not act as a spring force, which explains why this limit of validation is not observed.

%%%%%%%%%%%%%%%%%%%%%%%%%%%%%%%%%%%%%%%%%%%%%%%%%%%%%%%%%%%%%%%%%%%%%%%%%%%%%%%%%%%%%%%%%%%%%%%%%%%%%%%%%%%%%%%%%%%%%%%%%%%
\section{Discussion}

\begin{figure}
\begin{center}
\includegraphics[width=85mm]{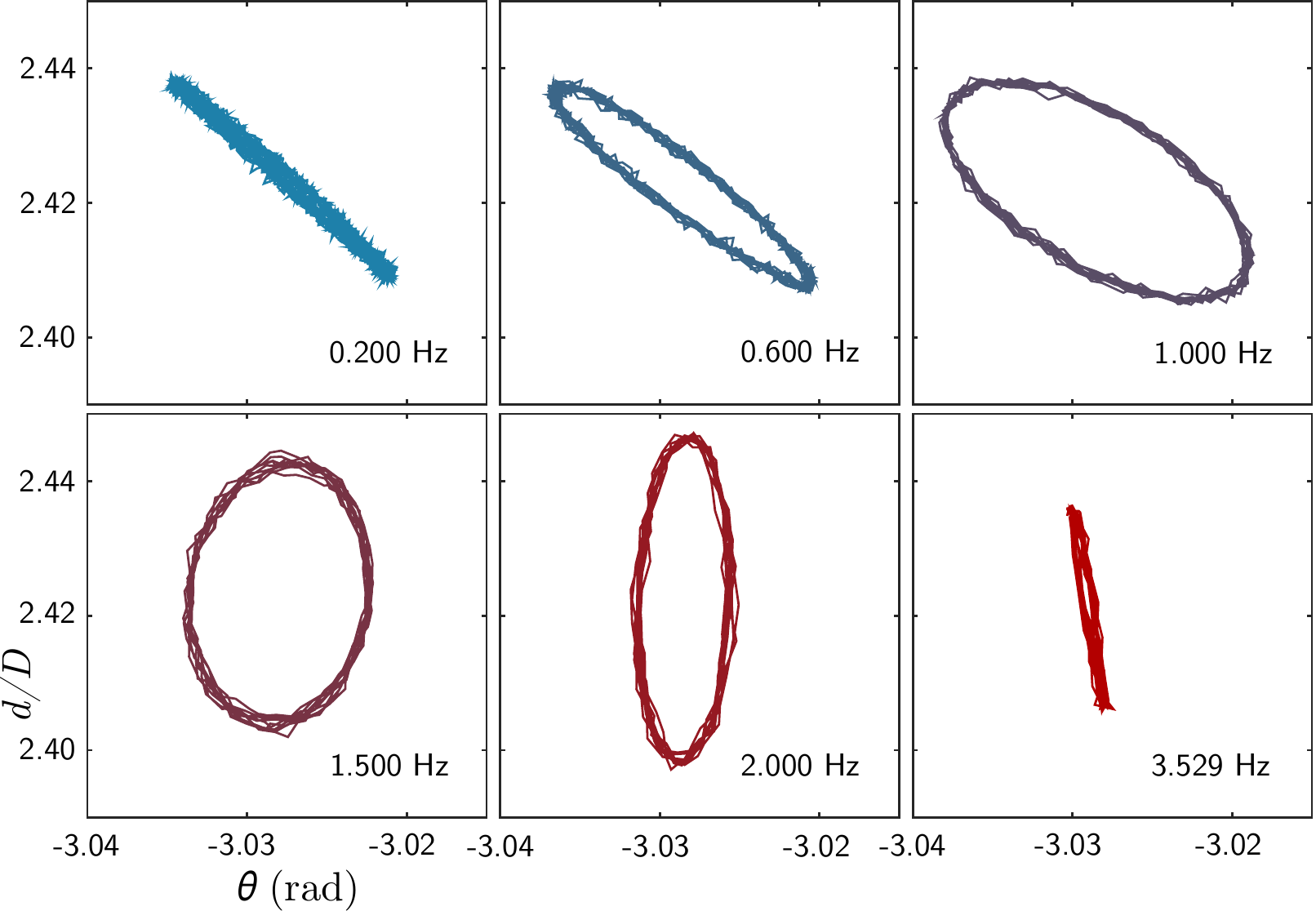}
\caption{
\textbf{Deformation cycles.} -- 
Trajectories in plane $\left( \theta, d / D \right)$ for 10 periods of the oscillating induction with $\alpha = 45^\circ$.
Below both resonance frequencies, in-phase oscillations produce no loop.
When the frequency is increased, loops appear as oscillations get out of phase.
Above both resonances, cycles disappear as phase difference decreases to zero.
}
\label{FIG5}
\end{center}
\end{figure}
Controlling the phase difference between degrees of freedom is the key ingredient for generating significant non-reciprocal deformations, which are necessary to low Reynolds propulsion~\cite{lauga2009}.
Indeed, to propel itself, a body immersed in a fluid must produce a net flow over a time period. This means that low Reynolds swimmers must find a way around the time-reversal properties of Stokes flows, either by using the medium to break symmetry~\cite{tierno2008} or by undergoing non-reciprocal, \emph{i.e.} non-time-reversible, deformations~\cite{lumay2013, lauga2009}. Such deformations can be obtained in a magnetocapillary assembly by applying time-dependent magnetic induction fields~\cite{lumay2013}. Note that the field itself can be time-reversible, as the system spontaneously breaks time-reversal symmetry through magnetocapillary interactions~\cite{grosjean2015}.

In this work, we evidenced that for magnetocapillary bonds, even if the fluid is in a Stokes regime, the Reynolds number is not the only parameter driving the dynamics of this system, the system inertia has to be taken into account.
Indeed, quality factor $Q=\omega_0 \tau /2$ can be over unity, meaning that the system is underdamped, granted that the stiffness of the potential and the particle mass are high enough.
This is of interest for creating phase differences between stretching and swinging of the bonds.
Each mode possesses a quality factor $Q$ which gives the shape of the resonance curve, and the ratio of the frequency to the resonance frequency $\xi=\omega/\omega_0$ allows to distinguish between the well known regimes of a forced harmonic oscillator.
Firstly, assuming $Q>1$ (for $Q<1$, the same discussion can be made using $1/Q$ instead of $Q$), if $\xi<1/Q$ the oscillator follows the forcing in a quasistatic way, in phase.
Then, if $Q>\xi>1/Q$, the dynamics is limited by the viscous dissipation and the oscillation of the system is in quadrature with the forcing.
Finally, if $\xi>Q$ the oscillator is limited by inertia and oscillates in phase opposition with the forcing.

The stretching and swinging modes of the magnetocapillary bond are generally detuned ($f_d \neq f_\theta$).
They can be both excited simultaneously using intermediate angles $\alpha$ ($0$ and $90^\circ$ produce pure radial and orthoradial oscillations respectively).
If the resonances are sufficiently separated i.e, $f_d/f_\theta<1/Q$ or $f_d/f_\theta>Q$, the excitation frequency can be chosen so that one mode oscillates in quadrature and the other oscillates in phase or phase opposition with the excitation as was shown in fig.~\ref{FIG3}.
Therefore, the system can follow simple non-reciprocal Lissajous curves (ellipses) in the plane $\left( \theta, d / D \right)$ whose shape depends on control parameters $f$ and $\alpha$.
Figure~\ref{FIG5} represents six cycles for $\alpha = 45^\circ$.
For each cycle, 10 periods of the oscillating induction are represented.
For frequencies below both resonances, \emph{i.e.} $f < f_d$ and $f < f_\theta$, in-phase oscillations produce reciprocal deformations.
When frequency is increased, cycles appear as oscillations get out of phase.
Around resonance frequencies, phase difference is close to $\pi / 2$.
At frequencies higher than both resonances, \emph{i.e.} $f > f_d$ and $f > f_\theta$, cycles disappear as phase difference decreases to zero.

Further downscaling of the system could prove useful for future applications, like low Reynolds swimming.
To determine if a phase difference is achievable at smaller scale, we can look at the expressions of quality factors in radial and orthoradial directions.
For a constant ratio $D/d$, we have
\begin{equation}
Q_\theta = \frac{\omega_\theta\tau}{2} \sim DB.
\end{equation}
This means that orthoradial q factor can be kept constant, granted that induction field $B$ is increased accordingly.
As a matter of fact, at the center of a Helmholtz system, we have $B\sim 1/R$ where $R$ is the radius of the coils.
If coil size is decreased along with particle diameter $D$, with everything else kept constant, the orthoradial oscillator will behave similarly.
In radial direction, however, we have
\begin{equation}
Q_d = \frac{\omega_d \tau}{2} \sim \frac{q D^2}{d} \sim \frac{D^5}{d}
\end{equation}
since capillary charge $q \sim D^3$. With a constant ratio $D/d$, we have $Q_d \sim D^4$ meaning that radial oscillation is overdamped at smaller scales.
As a result, for small scale systems, the only way to achieve non-reciprocal cycles is to use the orthoradial resonance frequency as excitation frequency. 

Of course, exciting radial and orthoradial modes using independent, out of phase magnetic induction fields would also generate non-reciprocal motion.
However, magnetocapillary bonds in assemblies of more than 2 particles are usually not aligned, meaning that both modes are always excited.
This may explain why a single oscillating sinusoidal field can generate efficient locomotion~\cite{lumay2013, grosjean2015}.
Previous theoretical work on magnetocapillary systems~\cite{chinomona2015} studied the Stokesian motion of particles in a magnetocapillary well, finding insufficient symmetry breaking to account for efficient low Reynolds locomotion in a three beads system.
However, the frequency was not a relevant parameter in their work.
Lastly, it is obvious from eq.~(\ref{U}) that for larger deformations, the potential becomes anharmonic.
Stretching and swinging degrees of freedom become coupled, opening a rich variety of nonlinear behaviour.

%%%%%%%%%%%%%%%%%%%%%%%%%%%%%%%%%%%%%%%%%%%%%%%%%%%%%%%%%%%%%%%%%%%%%%%%%%%%%%%%%%%%%%%%%%%%%%%%%%%%%%%%%%%%%%%%%%%%%%%%%%%
\section{Conclusion}

We demonstrated that magnetocapillary bonds can be created along liquid interfaces. 
Magnetocapillary numbers $M_d$ and $M_\theta$ determine the interdistance and orientation of the bond, respectively.
Then, we proved theoretically and experimentally that such bonds possess two vibration modes, a radial stretching mode and an orthoradial swinging mode, that can be excited independently.
These modes exhibit a resonance when excited, even though the system is in Stokes regime.
This means that bead inertia must be taken into account and that frequency is an essential control parameter of the bond.
Indeed, depending on which frequency is applied to the system, it can experience out of phase radial and orthoradial deformations.
This allows to generate a breaking of time reversibility, which is required for low Reynolds locomotion.
If this system were to be downscaled, similar dynamics would occur at the orthoradial resonance frequency.

While this paper focused on near equilibrium dynamics, nonlinear behaviours and coupling certainly matter for low Reynolds propulsion.
Studying the response of magnetocapillary bonds to finite perturbations would be the next logical step.
Further work should also determine how to combine several pairs of particles in larger systems.
Indeed, magnetocapillary bonds constitute versatile building blocks that can be combined to create elaborated microstructures.

%%%%%%%%%%%%%%%%%%%%%%%%%%%%%%%%%%%%%%%%%%%%%%%%%%%%%%%%%%%%%%%%%%%%%%%%%%%%%%%%%%%%%%%%%%%%%%%%%%%%%%%%%%%%%%%%%%%%%%%%%%%
\acknowledgments
This work was financially supported by the FNRS (Grant PDR T.0043.14) and by the University of Li\`ege (Grant
FSRC 11/36). GG thanks FRIA for financial support. GLa was financed by the University of Li\`ege and the European
Union through MSCA-COFUND-BeIPD project.

\bibliography{biblio}{}
\bibliographystyle{apsrev4-1}

\end{document}